\documentclass[aps,prd,twocolumn,amsmath,groupedaddress,amssymb,showpacs,floatfix,nofootinbib,longbibliography]{revtex4-1}
\pdfoutput=1
\usepackage{graphicx}
\usepackage{bm}
\usepackage{times}
\usepackage{hyperref}
\usepackage{slashed}
\usepackage{color}
\usepackage{aas_macros}
\usepackage{slashed}
\usepackage{lipsum}
\usepackage{subfigure}
\usepackage{multirow}
\usepackage{amsmath}
\usepackage{array} 
\usepackage{varwidth} 
\usepackage{enumitem}

\usepackage{scalerel}
\usepackage{tikz}
\usetikzlibrary{svg.path}

\definecolor{orcidlogocol}{HTML}{A6CE39}
\tikzset{
	orcidlogo/.pic={
		\fill[orcidlogocol] svg{M256,128c0,70.7-57.3,128-128,128C57.3,256,0,198.7,0,128C0,57.3,57.3,0,128,0C198.7,0,256,57.3,256,128z};
		\fill[white] svg{M86.3,186.2H70.9V79.1h15.4v48.4V186.2z}
		svg{M108.9,79.1h41.6c39.6,0,57,28.3,57,53.6c0,27.5-21.5,53.6-56.8,53.6h-41.8V79.1z M124.3,172.4h24.5c34.9,0,42.9-26.5,42.9-39.7c0-21.5-13.7-39.7-43.7-39.7h-23.7V172.4z}
		svg{M88.7,56.8c0,5.5-4.5,10.1-10.1,10.1c-5.6,0-10.1-4.6-10.1-10.1c0-5.6,4.5-10.1,10.1-10.1C84.2,46.7,88.7,51.3,88.7,56.8z};
	}
}

\newcommand\orcidicon[1]{\href{https://orcid.org/#1}{\mbox{\scalerel*{
				\begin{tikzpicture}[yscale=-1,transform shape]
				\pic{orcidlogo};
				\end{tikzpicture}
			}{1}}}}

\newcommand{\diff}{\mathrm{d}}
\newcommand{\e}{\mathrm{e}}

\newcommand{\sigmav}{\langle \sigma v \rangle}

\newcommand{\Td}{T_\mathrm{d}}
\newcommand{\ad}{a_\mathrm{d}}
\newcommand{\Tf}{T_\mathrm{f}}
\newcommand{\kfs}{k_\mathrm{fs}}
\newcommand{\Eann}{E_\mathrm{ann}}
\newcommand{\ac}{a_\mathrm{c}}
\newcommand{\Rp}{q}
\newcommand{\rhoC}{\bar\rho_0}
\newcommand{\Rc}{r_\mathrm{cusp}}
\newcommand{\rc}{r_\mathrm{core}}
\newcommand{\fmax}{f_\mathrm{max}}
\newcommand{\dc}{\delta_\mathrm{c}}

\begin{document}

\title{Can prompt cusps of WIMP dark matter be detected as individual gamma-ray sources?}
\author{M. Sten Delos \orcidicon{0000-0003-3808-5321}}
\email{mdelos@carnegiescience.edu}
\affiliation{The Observatories of the Carnegie Institution for Science, 813 Santa Barbara Street, Pasadena, CA 91101, USA}

\begin{abstract}
Prompt $\rho\propto r^{-1.5}$ density cusps are the densest and most abundant dark matter systems. If the dark matter is a weakly interacting massive particle (WIMP), recent studies have shown that prompt cusps dominate the aggregate dark matter annihilation rate. This article explores whether individual prompt cusps could be detected as gamma-ray sources. At the Fermi telescope's point-source sensitivity, WIMPs with the canonical annihilation cross section could form detectable prompt cusps if the particle mass is of order 10 GeV. These objects could be 10-100 pc away and weigh under a solar mass; they would subtend around 0.1 degrees on the sky. For GeV-scale dark matter particles with below-canonical cross sections, searches for individual prompt cusps can be more sensitive than searches for the annihilation signals from galactic dark matter halos.
\end{abstract}
\maketitle

\section{Introduction}

The earliest stages of structure formation in the Universe yield abundant $\rho\propto r^{-1.5}$ density cusps of dark matter \cite{2010ApJ...723L.195I,2013JCAP...04..009A,2014ApJ...788...27I,2015MNRAS.450.2172P,2016MNRAS.461.3385O,2017MNRAS.471.4687A,2018PhRvD..97d1303D,2018PhRvD..98f3527D,2018MNRAS.473.4339O,2019PhRvD.100b3523D,2020MNRAS.492.3662I,2021A&A...647A..66C,2022MNRAS.517L..46W,2023MNRAS.518.3509D,2023PDU....4101259D,2023arXiv230905707O}. These \textit{prompt cusps} form at the moment of collapse of smooth peaks in the initial density field, and their abundance and properties are closely related to those of the peaks \cite{2019PhRvD.100b3523D,2023MNRAS.518.3509D,2023arXiv230905707O}. Prompt cusps are expected to largely survive up to the present day \cite{2023MNRAS.518.3509D,2023JCAP...10..008D}. They are the densest dark matter structures.

Weakly interacting massive particles (WIMPs) are among the most prominent candidates for dark matter \cite{2018RPPh...81f6201R,2018EPJC...78..203A}. WIMPs are thermally produced in the early universe, and this process requires (under standard cosmological assumptions) that these particles have an annihilation cross section of about $\sigmav\simeq 2\times 10^{-26}$~cm$^3$s$^{-1}$ to yield the measured abundance of dark matter \cite{2012PhRvD..86b3506S}. But if the dark matter can annihilate, then the great abundance and internal density of prompt cusps make them dominate the annihilation rate \cite{2023JCAP...10..008D}. Annihilation within unresolved prompt cusps should produce a significant fraction of the isotropic gamma-ray background within a certain energy range \cite{2023JCAP...10..008D,2023MNRAS.523.1067S}, and nondetection of such a contribution implies constraints on WIMP dark matter that can be stronger than those obtained by searching for annihilation signals associated with galactic dark matter halos \cite{2023arXiv230713023D}.

Given their dominant contribution to the total annihilation rate, it is natural to ask whether prompt cusps could be resolved as individual sources of annihilation radiation. 
Prospects for individual detection of the smallest dark matter objects through their annihilation radiation were previously explored by Refs.~\cite{2005ApJ...633L..65O,2010ApJ...723L.195I,2006PhRvL..97s1301K,2008PhRvD..78j1301A,2009JCAP...07..007L,2014GrCo...20...47B}, and Refs.~\cite{2015JCAP...12..035B,2016JCAP...05..049B,2019JCAP...11..045C,2020PhRvD.102j3010D,2021ApJ...906...57S,2021PDU....3200845C,2023MNRAS.520.1348G} have searched for annihilation radiation from more massive dark matter systems that are still too small to host galaxies.
However, the recently developed prompt-cusp paradigm has greatly clarified the abundance, properties, and survival of the smallest dark matter systems.
In this article, we use recent characterizations of the prompt cusp population \cite{2019PhRvD.100b3523D,2023MNRAS.518.3509D,2023JCAP...10..008D} and the degree to which they are disrupted in our own vicinity within the Galaxy \cite{2023MNRAS.523.1067S,2023MNRAS.tmp..828S} to explore detection prospects for individual prompt cusps.

We consider WIMP models with a range of masses $m$ and kinetic decoupling temperatures $\Td$. Generally, lower $m$ and $\Td$ yield fewer and more massive prompt cusps, so detection prospects for individual cusps in these models are better. However, at the Fermi telescope's point-source sensitivity of $\sim 10^{-12}$~erg~cm$^{-2}$s$^{-1}$ \cite{2020ApJS..247...33A} and assuming a canonical thermal-relic cross section, only dark matter models with masses smaller than $\mathcal{O}(10)$~GeV (depending on $\Td$) yield detectable prompt cusps. Such masses are likely already excluded by other observations (e.g.~\cite{2016PhRvD..93b3527S,2017ApJ...834..110A,2023arXiv230713023D}). Point-source sensitivity would need to be better by at least an order of magnitude in order to possibly detect prompt cusps in still-viable WIMP models under standard cosmological assumptions. However, for GeV-scale models with below-canonical cross sections, searches for individual prompt cusps can be competitive with other strategies.

This article is organized as follows. Section~\ref{sec:dist} describes how we characterize the local population of prompt cusps. Section~\ref{sec:flux} arrives at the main results: the expected flux of the brightest prompt cusp, for a range of WIMP scenarios, and limits on the annihilation cross section that arise therefrom. Section~\ref{sec:props} explores properties of the brightest prompt cusps, including their angular sizes. Finally, we conclude in Sec.~\ref{sec:conclusion}.

\section{Characterizing the prompt cusp population}
\label{sec:dist}

Prompt cusps arise from the collapse of peaks in the initial density field, and we follow the procedure of Ref.~\cite{2023JCAP...10..008D} to determine the distribution of cusps from the distribution of peaks.

\subsection{Dark matter power spectrum}\label{sec:power}

\begin{figure}[tbp]
	\centering
	\includegraphics[width=\columnwidth]{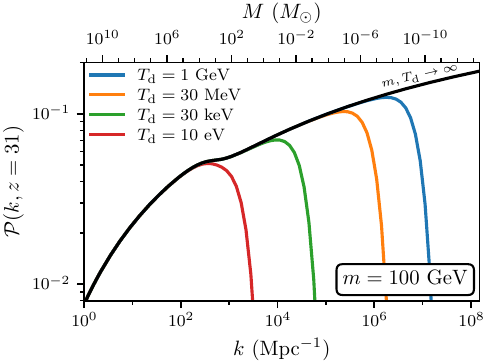}
	\caption{Dimensionless power spectrum $\mathcal{P}(k)\equiv [k^3/(2\pi^2)]P(k)$ of dark matter density variations, evaluated at linear order at $z=31$. The black curve represents arbitrarily cold dark matter, while the colored curves represent WIMPs with mass $m=100$~GeV that kinetically decouple at a range of temperatures $\Td$.}
	\label{fig:power}
\end{figure}

The peak distribution is set by the linear dark matter power spectrum, which we evaluate at redshift $z=31$ using the \textsc{class} code \cite{2011JCAP...07..034B} and extrapolate below the code's resolution limit using the analytic small-scale growth function in Ref.~\cite{1996ApJ...471..542H}. We adopt cosmological parameters as measured by the Planck mission \cite{2020A&A...641A...6P}. This power spectrum is shown as the black curve in Fig.~\ref{fig:power} and describes dark matter of arbitrarily low temperature. The kink around $k\sim 10^{2.5}$~Mpc$^{-1}$ arises because baryons cluster with the dark matter at larger scales (smaller $k$) but resist clustering on smaller scales (larger $k$) \cite{2006PhRvD..74f3509B}. Dark matter density contrasts $\delta\equiv(\rho-\bar\rho)/\bar\rho$ grow as $\delta\propto a$ in the former case and $\delta\propto a^g$ with $g\simeq 0.901$ in the latter case \cite{1996ApJ...471..542H}, where $a$ is the scale factor.

WIMPs are produced from the hot primordial plasma and retain some thermal motion as a result, which smooths out density variations on a characteristic free-streaming scale. We assume that the dark matter kinetically decouples from the plasma at some temperature $\Td$, which could be lower (but not higher) than the temperature $\Tf\simeq m/20$ at which the dark matter chemically decouples and its abundance freezes out. Here $m$ is the mass of the dark matter particle. We assume that free streaming multiplies the power spectrum by $\exp(-k^2/\kfs^2)$, consistently with a Maxwellian velocity distribution, with a free-streaming wavenumber $\kfs$ evaluated as described in Ref.~\cite{2006PhRvD..74f3509B}. The colored curves in Fig.~\ref{fig:power} represent the linear dark matter power spectra at $z=31$ for WIMPs of mass $m=100$~GeV with a range of decoupling temperatures $\Td$. Since the dark matter is maintained at the temperature of the plasma until temperature $\Td$, higher $\Td$ leads to colder dark matter and a shorter free-streaming scale.

Since prompt cusps form from smooth peaks in the density field, they arise at the smoothing scale, i.e., the scale at which free streaming starts to significantly suppress the power spectrum. As a guide to the typical masses of these objects, the upper axis in Fig.~\ref{fig:power} shows the mass scale $M\equiv (4\pi/3)\rhoC k^{-3}$ associated with each wavenumber $k$, where $\rhoC$ is the comoving dark matter density. For the range of WIMP scenarios shown, the free-streaming mass scale ranges from about $10^{-8}$~M$_\odot$ to about $10^3$~M$_\odot$.

\subsection{Density peaks and prompt cusps}\label{sec:peaks}

Prompt cusps are tightly related to the density peaks that form them. A peak that collapses at the scale factor $\ac$ yields a prompt cusp with density profile $\rho=A r^{-1.5}$, where
\begin{equation}
    A \simeq 24 \rhoC \ac^{-1.5} \Rp^{1.5}
\end{equation}
\cite{2019PhRvD.100b3523D,2023MNRAS.518.3509D}.
Here $\Rp\equiv|\delta/\nabla^2\delta|^{1/2}$ is the characteristic comoving size of the peak, defined in terms of its height $\delta\equiv(\rho-\bar\rho)/\bar\rho$ and comoving curvature $\nabla^2\delta$. This profile extends out to at least the radius
\begin{equation}
    \Rc\simeq 0.11 \ac\Rp
\end{equation}
\cite{2023MNRAS.518.3509D}. There is also an inner radius
\begin{equation}
    \rc\simeq 0.1 G^{-2/3}\fmax^{-4/9}A^{-2/9}
\end{equation}
below which the $\rho\propto r^{-3/2}$ cusp must give way to a finite-density core \cite{2023MNRAS.518.3509D}; it is related to the maximum phase-space density $\fmax=(2\pi)^{-3/2}(m/\Td)^{3/2}\rhoC\ad^{-3}$ set in the early Universe, where $\ad$ is the scale factor of kinetic decoupling (when the temperature is $\Td$).

Given the power spectrum, Ref.~\cite{1986ApJ...304...15B} derived the number density of peaks in the density field and their distribution in height $\delta$ and curvature $\nabla^2\delta$, whence the characteristic size $q$ follows directly. The collapse time $\ac$ of a peak of height $\delta$ is given by $[D(\ac)/D(a)] \delta = \dc(e,p)$, where $D$ is the linear growth function, $\dc$ is the linear threshold for ellipsoidal collapse, and $a=1/32$ is the scale factor at which the power spectrum was evaluated. We evaluate $\dc(e,p)$ using the approximation in Ref.~\cite{2001MNRAS.323....1S}; the distribution of the ellipticity $e$ and prolateness $p$ of the tidal field is also given in Ref.~\cite{2001MNRAS.323....1S}. For the growth function, we approximate $D(a)=a^g$ with $g\simeq 0.901$, which is valid for dark matter at wavenumbers $k\gtrsim 10^3$~Mpc$^{-1}$ (see Sec.~\ref{sec:power}). Since our analysis includes slightly larger scales as well (see Fig.~\ref{fig:power}), this growth function is not completely accurate. However, since the $z=31$ evaluation time of the linear matter power spectrum is already close to the collapse times of the peaks whose prompt cusps dominate the annihilation signal \cite{2023JCAP...10..008D}, any error resulting from using a slightly incorrect growth function to evaluate collapse times is expected to be minimal.

\begin{figure}[tbp]
	\centering
	\includegraphics[width=\columnwidth]{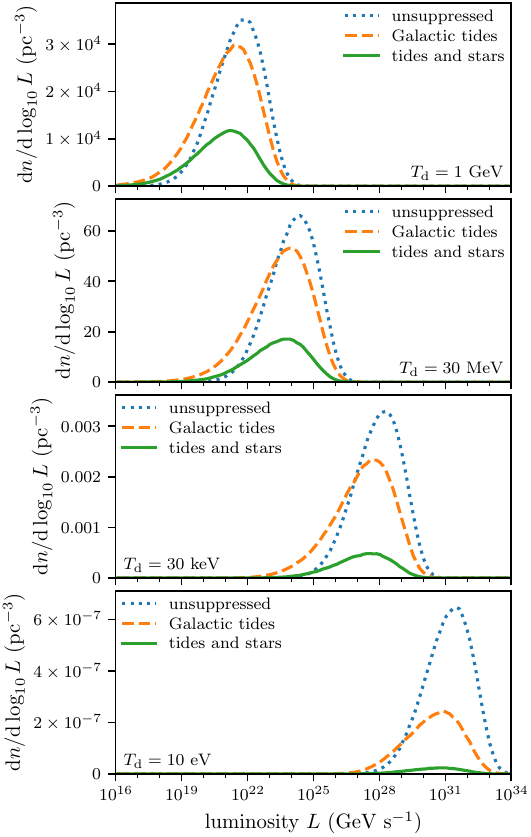}
	\caption{Local number density of prompt cusps per decade in gamma-ray luminosity $L$ for 100-GeV WIMPs with the canonical annihilation cross section. Different panels consider different decoupling temperatures $\Td$. The dotted curves assume that prompt cusps survive fully intact, while the solid curves account for disruption of prompt cusps by tidal forces and encounters with stars. The dashed curves account for tidal forces but not stellar encounters. As Sec.~\ref{sec:local} discusses, the stellar encounter modeling used to make the solid curves becomes inaccurate for high prompt cusp masses (corresponding to low $\Td$), and stellar encounters actually become a negligible consideration in this regime.}
	\label{fig:distL}
\end{figure}

For each dark matter model, this calculation yields the cosmologically averaged abundance and distribution of prompt cusps. We scale the abundance by the approximate factor $1/2$ to account for loss of prompt cusps during the hierarchical growth of structure in the universe \cite{2023JCAP...10..008D}. For the distribution, we Monte Carlo sample about $10^6$ prompt cusps for each scenario, each of which has a density coefficient $A=\rho r^{1.5}$, outer radius $\Rc$, and core radius $\rc$.
Assuming the dark matter self-annihilates with a velocity-independent cross section $\sigmav$,\footnote{Prompt cusps have the lowest internal velocity dispersion of all collapsed dark matter systems (potentially of order m/s), so they are not a significant source of annihilation radiation in models where $\sigma v$ scales as a positive power of $v$ at lowest order.}
the annihilation rate inside a prompt cusp is
\begin{equation}\label{Gamma}
    \Gamma = \frac{\sigmav}{2m^2}\int\rho^2\diff V,
\end{equation}
where we integrate the squared density $\rho^2$ over the volume of the cusp. For consistency with later analyses, we assume prompt cusps have the density profile in Ref.~\cite{2023MNRAS.523.1067S}, which describes a $\rho=Ar^{-1.5}$ cusp modified in phase space to not exceed the maximum phase-space density $\fmax$. For this density profile,
\begin{equation}
    \int\rho^2\diff V\simeq 4\pi A^2 [0.531+\log(\Rc/\rc)].
\end{equation}
The dotted curves in Fig.~\ref{fig:distL} show, for 100~GeV WIMPs with several values of $\Td$, the distribution of prompt cusps in terms of their annihilation luminosity
\begin{equation}\label{L}
    L=\Eann \Gamma,
\end{equation}
where $\Eann$ is the energy released in gamma rays per annihilation.
We assume for simplicity that $\Eann=m/2$, i.e., that a quarter of the annihilation energy is released as gamma radiation. This is typical for heavy quark or boson channels, such as $b\bar b$ or $W^+W^-$; it is a moderate overestimate for $\tau^+\tau^-$. We also assume that $\sigmav=2.2\times 10^{-26}$~cm$^3$s$^{-1}$, as required for the dark matter to reach the known abundance.

\subsection{Local prompt cusps}\label{sec:local}

Observationally detectable prompt cusps would be quite close to the Sun (we will quantify precisely how close in Sec.~\ref{sec:props}).
The local dark matter density is about $2.7\times 10^5$ times higher than the cosmological average \cite{2020MNRAS.494.4291C}, so the local number density of prompt cusps is also higher by the same factor. This scaling sets the normalization of the differential number density $\diff n/\diff\log_{10}L$ of prompt cusps that is shown as the dotted curves in Fig.~\ref{fig:distL}. The general trend of this figure is that smaller $\Td$ (later decoupling) leads to more massive prompt cusps (compare Fig.~\ref{fig:power}), which are more luminous (higher $L$) but correspondingly less abundant (lower $n$).

Prompt cusps within the Galaxy are also altered by tidal stripping and encounters with stars. We use the \textsc{cusp-encounters} code \cite{2023MNRAS.523.1067S} to model these effects. This code adopts an observationally motivated description of the Galactic dark matter halo and star distribution, integrates randomly sampled prompt cusp orbits within the Galactic potential, and models the impact of tidal forces \cite{2023MNRAS.tmp..828S} and encounters with stars \cite{2023MNRAS.523.1067S} along that orbit. We restrict the sample of orbits to those that currently reside at the Sun's Galactocentric radius of 8.2~kpc. The solid curves in Fig.~\ref{fig:distL} show the differential number density $\diff n/\diff\log_{10}L$ of prompt cusps when Galactic tidal forces and stellar encounters are accounted for. The dashed curves show the influence of tidal forces alone. Generally, these effects reduce the abundance of prompt cusps at given $L$ by a factor of a few up to around an order of magnitude. Lower $\Td$ yields prompt cusps of lower internal density, which are more susceptible to disruption.

However, the \textsc{cusp-encounters} model overestimates the impact of stellar encounters for models with particularly low $\Td$ ($\lesssim 10^{-4}$~GeV). The reason is that the influence of stars that pass through a cusp is much lower than would be predicted by the ``distant tide'' approximation made by Ref.~\cite{2023MNRAS.523.1067S}, which considers the star to be arbitrarily far away \cite{1993ApJ...413L..93M,2019PhRvD.100h3529D,2022arXiv220109788F}.\footnote{Under the distant tide approximation, the energy injected into a cusp by a stellar encounter scales as $b^{-4}$, where $b$ is the impact parameter. Reference~\cite{1993ApJ...413L..93M} found in simulations of globular clusters, and Ref.~\cite{2019PhRvD.100h3529D} found in simulations of dark matter halos, that a $(b^4+r^4)^{-1}$ scaling is more appropriate, where $r$ is a measure of the size of the cluster or halo.} Models with low $\Td$ yield particularly massive prompt cusps, which are large enough that these penetrative stellar encounters are common. As a very approximate treatment, we check for each dark matter model whether, for the brightest cusps, the most disruptive stellar encounter passed closer than the radius at which that encounter is predicted to truncate the cusp.\footnote{\label{foot:testb} We take the truncation radius to be $0.4r_B$, where $r_B$ is the characteristic radius associated with the stellar encounter, defined in Ref.~\cite{2023MNRAS.523.1067S} as a function of the tidal impulse $B$. $0.4r_B$ is roughly the radius at which the encounter suppresses the cusp's density profile by a factor of 2. 
We compare this to the impact parameter $b$, but for efficiency reasons, the \textsc{cusp-encounters} code does not sample $b$ directly. As another approximation, we reconstruct $b$ from the tidal impulse $B$ by assuming a solar-mass star that passes at 200~km~s$^{-1}$. We average the resulting ratio $b/(0.4r_B)$ over the cusp distribution at fixed brightness, as described in Sec.~\ref{sec:props}.} If the encounter distance is smaller than the truncation radius, then we simply neglect stellar encounters. The end result is that stellar encounters are neglected for cusps more massive than about $10^{-2}$~M$_\odot$; this mass threshold is approximately in agreement with the predictions in Ref.~\cite{2022arXiv220109788F}.

\section{Brightness of prompt cusps}\label{sec:flux}

We now evaluate the distribution of the brightness of prompt cusps as seen from the Earth. The energy flux from a source with luminosity $L$ at distance $d$ is $F=L/(4\pi d^2)$.
We can find the expected number of sources with flux exceeding $F$ by integrating the differential cusp number density in Fig.~\ref{fig:distL} over $L$ and volume subject to the flux constraint, i.e.,
\begin{align}
    N(>F) &= \int_{L/(4\pi d^2)>F} 4\pi d^2\diff d \diff L \frac{\diff n}{\diff L}.
\end{align}
By substituting $d=\sqrt{L/(4\pi F^\prime)}$ and integrating over $L$ and $F^\prime>F$, we obtain
\begin{align}\label{NF0}
    N(>F) &=
    \frac{1}{2\sqrt{4\pi}}
    \int_F^\infty\frac{\diff F^\prime}{F^{\prime 5/2} }
    \int_0^\infty\diff L\, L^{3/2} \frac{\diff n}{\diff L}
    \\\label{NF}
    &=(F_*/F)^{3/2},
\end{align}
where
\begin{align}
    F_*\equiv \left(\frac{1}{3\sqrt{4\pi}}\int_0^\infty \diff L\, L^{3/2} \frac{\diff n}{\diff L}\right)^{2/3}.
\end{align}
Note that $F_*$ is the flux threshold above which one source is expected. From Poisson statistics, the probability that at least one source exceeds flux $F$ is
\begin{equation}\label{PF}
    P(>F)=1-\e^{-N(>F)}=1-\e^{-(F_*/F)^{3/2}}.
\end{equation}
Inverting this expression yields that with probability $P$, at least one source exceeds the flux 
\begin{equation}\label{Fprob}
    F_P=[-\log(1-P)]^{-2/3}F_*.
\end{equation}
For example, the median flux of the brightest source is $F_{0.5}\simeq 1.28F_*$, and in 95 percent of cases, at least one source is expected to exceed flux $F_{0.95}\simeq 0.48F_*$.

For a 100-GeV dark matter particle that decouples at 30~MeV, typical parameters for a WIMP model \cite{2004MNRAS.353L..23G}, $F_*\simeq 6 \times 10^{-15}$~erg~cm$^{-2}$~s$^{-1}$. Probing this scenario would require more than two orders of magnitude better sensitivity than the Fermi point-source sensitivity of $\sim 10^{-12}$~erg~cm$^{-2}$~s$^{-1}$ \cite{2020ApJS..247...33A}.\footnote{The Fermi point-source sensitivity limit is quoted as about $2\times 10^{-12}$ erg~cm$^{-2}$s$^{-1}$, but it is lower for the strongly curved spectra that would arise from dark matter annihilation.}

\begin{figure}[tbp]
	\centering
	\includegraphics[width=\columnwidth]{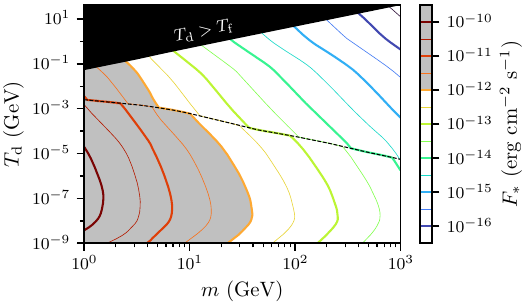}
	\caption{
        Characteristic flux $F_*$ that one prompt cusp is expected to exceed if the dark matter has mass $m$, kinetically decouples at temperature $\Td$, and has the canonical thermal-relic cross section \cite{2012PhRvD..86b3506S}. The median flux from the brightest cusp is about $1.3F_*$, and at 95 percent confidence, at least one cusp exceeds a flux of about $F_*/2$. Low dark matter masses and decoupling temperatures yield prompt cusps that are individually more massive, leading to higher $F_*$. The shaded region marks $F_*>10^{-12}$~erg~cm$^{-2}$s$^{-1}$, approximately the regime accessible to the Fermi telescope \cite{2020ApJS..247...33A}.
        The black region is physically nonsensical because it would imply that the dark matter kinetically decoupled before it was produced by freeze-out.
        Above the dashed line, cusps are disrupted by stellar encounters, while below it, stellar disruption is neglected (see Sec.~\ref{sec:local}); in reality the transition between the regimes should be gradual.
        }
	\label{fig:Fstar}
\end{figure}

Figure~\ref{fig:Fstar} shows the broader picture of $F_*$ as a function of $m$ and $\Td$. We assume that for each particle mass $m$, $\sigmav$ takes the value needed to achieve the correct abundance through freeze-out, as calculated by Ref.~\cite{2012PhRvD..86b3506S}. We leave out the region for which kinetic decoupling occurs before freeze-out (black triangle), since this is physically nonsensical. The dashed line shows the threshold for stellar encounters to be accounted for (above the line) or neglected (below the line).

Generally, lower $m$ and lower $\Td$ lead to longer free-streaming scales and hence larger prompt cusps, which are more detectable on an individual level. We note in particular that $F_*\propto n^{2/3}\langle L^{3/2}\rangle^{2/3}$, where $n$ is the cusp number density and the angle brackets average over the cusp distribution. As cusp masses $M$ increase, it is approximately the case that $L\propto M\sigmav/m$ (where the $\sigmav/m$ factor comes from Eqs. \ref{Gamma} and~\ref{L}) and $n\propto 1/M$, so that $F_*\propto M^{1/3}\sigmav/m$. Noting that $M\propto \kfs^{-3}$ and $\kfs\propto m^{1/2}\Td^{1/2}$ \cite{2004MNRAS.353L..23G}, we obtain in the end that
\begin{equation}\label{scaling}
    F_*\propto m^{-3/2}\Td^{-1/2}\sigmav.
\end{equation}
This proportionality is only approximate and neglects logarithmic dependencies; it is only intended to supply intuition about the behavior of $F_*$.

For $\Td\lesssim 10^{-7}$~GeV, $F_*$ no longer increases with decreasing $\Td$ and instead decreases again. This effect arises because prompt cusps and their central cores become less dense as $\Td$ is decreased.  In the $\Td\lesssim 10^{-7}$~GeV regime, the central cores have low enough density that they begin to be significantly affected by tidal stripping in the Galactic potential. It is at this point that the annihilation rates in prompt cusps start to be greatly reduced by tidal stripping (see the dashed curve in the bottom panel of Fig.~\ref{fig:distL}). However, it should also be noted that very low decoupling temperatures mean that the dark matter has a large scattering cross section with baryons, and this can be ruled out by laboratory experiments (e.g.~\cite{2013PhRvD..88a5027C}).

The shaded region of Fig.~\ref{fig:Fstar} marks where $F_*$ exceeds the Fermi telescope's sensitivity of $\sim 10^{-12}$~erg~cm$^{-2}$s$^{-1}$ to point sources \cite{2020ApJS..247...33A}. Only dark matter models with masses of order 10~GeV would result in individually detectable prompt cusps. Masses below $\mathcal{O}(100)$~GeV are already likely ruled out by nondetection of the aggregate annihilation signal from prompt cusps \cite{2023arXiv230713023D} and by searches for annihilation radiation from galactic halos \cite{2017ApJ...834..110A}. Masses below about 20~GeV are also ruled out more robustly by the cosmic microwave background \cite{2016PhRvD..93b3527S}. Point-source sensitivity would need to better by one to three orders of magnitude, depending on $\Td$, in order for individual prompt cusps of WIMP dark matter with the canonical thermal-relic cross section to present useful targets for gamma-ray searches.

\begin{figure}[tbp]
	\centering
	\includegraphics[width=\columnwidth]{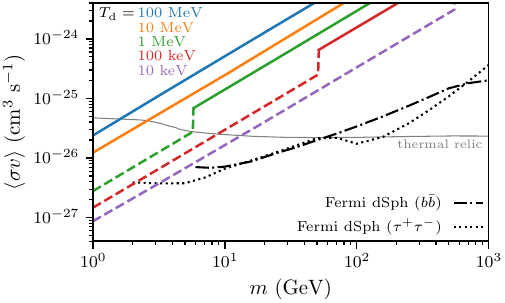}
	\caption{Limits on the annihilation cross section $\sigmav$ as a function of dark matter mass $m$. For several values of the decoupling temperature $\Td$, the colored lines show the 95 percent confidence limits, obtained under the hypothetical assumption that none of the Fermi-identified gamma-ray sources can be attributed to prompt cusps. These lines are solid in the regime where disruption of prompt cusps by stellar encounters is accounted for and dashed in the regime where it is neglected; in reality the transition should be more gradual. The thin gray line shows the cross section for dark matter produced by freeze-out under standard cosmological assumptions \cite{2012PhRvD..86b3506S}. For comparison, we also show limits on annihilation into $b\bar b$ (dot-dashed curve) and $\tau^+\tau^-$ (dotted curve) derived by searching for the signal from nearby dwarf galaxies \cite{2017ApJ...834..110A}. The limit from individual prompt cusps is weaker, but it can become competitive for low dark matter masses $m$ and decoupling temperatures $\Td$ (but note that the $b$ and $\tau$ masses are 4.2 and 1.8~GeV, respectively, so the fact that the black curves cut off near these masses is not reflective of the sensitivity of the associated search strategy).}
	\label{fig:limit}
\end{figure}

We can also relax the requirement that $\sigmav$ take its canonical thermal-relic value and test the limits on $\sigmav$ that would arise if none of the unidentified Fermi point sources \cite{2020ApJS..247...33A} could be attributed to dark matter annihilation. By requiring that $F_{0.95}\simeq 0.48F_*$ lie below the Fermi point-source sensitivity, we show upper limits on $\sigmav$ in Fig.~\ref{fig:limit} as a function of $m$ for several values of $\Td$ (different colors). The limits are depicted with solid lines when disruption by stellar encounters is included and with dashed lines when it is neglected (see Sec.~\ref{sec:local}). Our analysis is agnostic to the annihilation channel only because we neglect dependence of the sensitivity limit on the spectrum of the energy flux and we assume that a quarter of the annihilation energy is always released as gamma radiation. But for comparison, we show limits on $\sigmav$ for annihilation into $b\bar b$ and $\tau^+\tau^-$ derived from a search for the annihilation signal from the halos of nearby dwarf galaxies \cite{2017ApJ...834..110A}. These limits are mostly stronger than those from individual prompt cusps by a substantial margin, but they become comparable for dark matter masses $m\sim\mathcal{O}(1)$~GeV and low kinetic decoupling temperatures $\Td$.
However, we emphasize that the limits from individual cusps are hypothetical because although most of the unidentified point sources do not have spectral characteristics consistent with dark matter annihilation, some of them might \cite{2023MNRAS.520.1348G}.

\section{Properties of the brightest cusps}\label{sec:props}

We can also explore the properties of the brightest prompt cusps. From Eq.~(\ref{NF0}), the joint distribution of $F$ and $L$ is
\begin{equation}
    \frac{\diff^2 N}{\diff F\diff L}
    =\frac{1}{2\sqrt{4\pi}} F^{-5/2} L^{3/2} \frac{\diff n}{\diff L}.
\end{equation}
Thus, the distribution of prompt cusps at any fixed flux $F$ is just the underlying cusp distribution weighted by $L^{3/2}$.

\begin{figure}[tbp]
	\centering
	\includegraphics[width=\columnwidth]{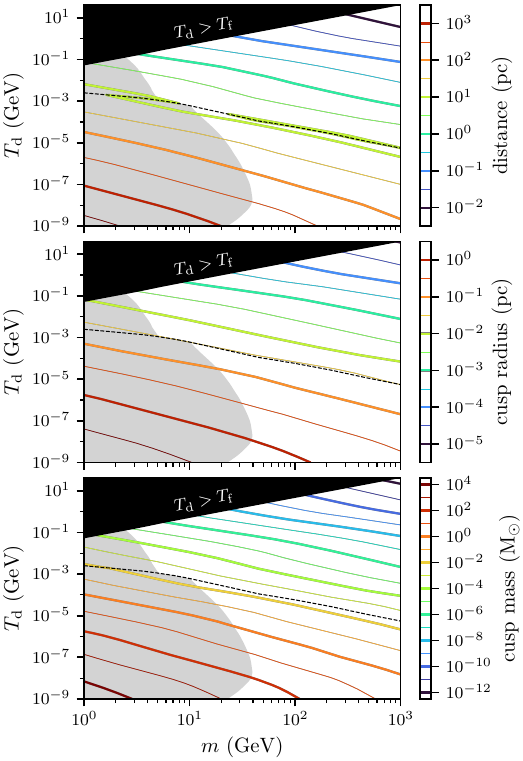}
	\caption{Properties of the brightest prompt cusps (as seen from Earth) for WIMP models with mass $m$ and kinetic decoupling temperature $\Td$. From the top, the three panels show the average distance, radius, and mass of cusps with flux $F_*$, respectively, where $F_*$ is the characteristic flux above which one cusp is expected (see Fig.~\ref{fig:Fstar}). As in Fig.~\ref{fig:Fstar}, the faintly shaded region is accessible given the Fermi telescope's point-source sensitivity of $\sim 10^{-12}$~erg~cm$^{-2}$s$^{-1}$ \cite{2020ApJS..247...33A}, the black region is excluded because it would require that the dark matter kinetically decouples before freeze-out, and disruption of cusps by stellar encounters is important above the dashed line and neglected below it.}
	\label{fig:stats}
\end{figure}

The upper panel of Fig.~\ref{fig:stats} shows the mean distance $d=\sqrt{L/(4\pi F_*)}$ of prompt cusps that produce flux $F_*$. The distance is generally very small in relation to Galactic scales, which justifies the approximation in Sec.~\ref{sec:local} that all cusps relevant to individual detection are at a Galactocentric radius of 8.2~kpc and the implicit assumption that the cusps are uniformly distributed in space. Only for very low decoupling temperatures $\Td\lesssim 10$~eV does the distance start to approach multiple kpc.

We next explore the sizes of the brightest prompt cusps. We take the radius $r$ of a cusp to be the smaller of its initial radius $\Rc$ and the radius at which stellar encounters and tidal forces truncate the density profile.\footnote{As in footnote~\ref{foot:testb}, we take the truncation radius to be $0.4r_B$, which is approximately the radius at which the density profile is suppressed by a factor of 2. However, here we evaluate $r_B$ (defined in Ref.~\cite{2023MNRAS.523.1067S}) based on the effective tidal impulse $B_{\mathrm{eff},\lambda}$ (defined in Ref.~\cite{2023MNRAS.523.1067S}) from Galactic tides and all stellar encounters combined.} The central panel of Fig.~\ref{fig:stats} shows for each WIMP model the average radius of cusps that produce a set energy flux at the Earth's position. For the models under consideration, the radius ranges from $10^{-5}$~pc up to several pc. Comparison with the upper panel reveals that $r/d\sim 2\times 10^{-3}$ in all cases, i.e., that the angular sizes of the brightest cusps are invariably about 0.1~deg. This is comparable to the width of the Fermi telescope's point-spread function at energies above about 10~GeV \cite{2013ApJ...765...54A}. However, even within this angular extent, the steep $\rho\propto r^{-1.5}$ density cusp produces a signal that is heavily concentrated toward the center. Thus, although searching for spatial extension is a common strategy to distinguish dark matter systems from astrophysical gamma-ray sources \cite{2015JCAP...12..035B,2016JCAP...05..049B,2019JCAP...11..045C,2020PhRvD.102j3010D}, this strategy may be difficult to employ for prompt cusps.

Finally, we explore the masses of the brightest prompt cusps, which we evaluate as $M=(8\pi/3)A r^{1.5}$, where $r$ is the radius considered above. The lower panel of Fig.~\ref{fig:stats} shows the average mass of cusps that produce a set energy flux at the Earth's position. These masses range from $\sim 10^{-12}$~M$_\odot$ for high $m$ and $\Td$ up to $\sim 10^4$~M$_\odot$ for low $m$ and $\Td$. Aside from being detectable as individual gamma-ray sources, such $\sim 10^4$~M$_\odot$ prompt cusps may even be susceptible to gravitational detection (e.g.~\cite{2023MNRAS.522L..78D}), although the models with very low $\Td$ are already challenged by terrestrial experiments \cite{2013PhRvD..88a5027C}. As we noted in Sec.~\ref{sec:local}, the threshold for the transition from stellar encounters being important to being negligible (dashed curve) corresponds to cusp masses of around $10^{-2}$~M$_\odot$.

\section{Conclusion}\label{sec:conclusion}

Prompt $\rho\propto r^{-1.5}$ density cusps arise in abundance at the onset of structure formation and largely survive up to the present time. In WIMP dark matter models, these objects dominate the aggregate dark matter annihilation rate. However, due to their low individual masses, prompt cusps are difficult to detect as individual sources of annihilation radiation.

Dark matter models with lower particle mass and that kinetically decouple from the plasma later tend to yield individually larger prompt cusps, which are more susceptible to individual detection as gamma-ray sources. At current levels of point-source sensitivity, and assuming a canonical thermal-relic cross section, prompt cusps of $\mathcal{O}(10)$~GeV dark matter could be detected. However, such models already tend to be ruled out by previous searches for annihilation radiation.

In order to detect prompt cusps in viable WIMP dark matter models with $\mathcal{O}(100)$~GeV mass and canonical cross section, sensitivity to point sources would need to be one to three orders of magnitude better than the Fermi telescope's point-source sensitivity limit of $\sim 10^{-12}$ erg~cm$^{-2}$s$^{-1}$, depending on when the dark matter kinetically decoupled. Alternatively, a low-mass WIMP with below-canonical cross section (which could be allowed by a nonstandard thermal history of the Universe \cite{2021OJAp....4E...1A}) can yield individually detectable prompt cusps at current sensitivity levels while not being already ruled out.
The annihilation signal from a prompt cusp is so spatially compact that consideration of spectral characteristics, such as the curvature of the energy spectrum \cite{2023MNRAS.520.1348G}, might be necessary to distinguish it from an astrophysical gamma-ray source. Another possibility is to search for a source with significant proper motion (e.g.~\cite{2006PhRvL..97s1301K}), since any detectable prompt cusp would lie at subkiloparsec distance.

\section*{Acknowledgements}

The author thanks Tim Linden and Daniel Ega{\~n}a-Ugrinovic for helpful discussions and comments on the manuscript.
This research was supported in part by Perimeter Institute for Theoretical Physics. Research at Perimeter Institute is supported by the Government of Canada through the Department of Innovation, Science and Economic Development and by the Province of Ontario through the Ministry of Research, Innovation and Science.

\bibliography{main}
\end{document}